\documentclass[aps,prb,twocolumn,floatfix,showpacs,10pt]{revtex4-1}
\usepackage[english]{babel}
\usepackage{amsmath,amssymb}
\usepackage[dvips]{graphicx}
\usepackage{color}
\begin{document}
\title{Plastic strain is a mixture of avalanches and \\ quasi-reversible
deformations: Study of various sizes}

\author{P\' eter Szab\' o}
\email{pszabo@metal.elte.hu}
\affiliation{Department of Materials Physics, E\" otv\" os University 
Budapest, H-1518 Budapest POB 32, Hungary}

\author{P\'eter Dus\'an Isp\'anovity}
\affiliation{Department of Materials Physics, E\" otv\" os University 
Budapest, H-1518 Budapest POB 32, Hungary}

\author{Istv\'an Groma}
\affiliation{Department of Materials Physics, E\" otv\" os University 
Budapest, H-1518 Budapest POB 32, Hungary}
 
\begin{abstract} 
Size-dependence of plastic flow is studied by discrete dislocation dynamical
simulation of systems with various numbers of interacting linear edge
dislocations while the stress is slowly increased.  Regions between avalanches
in the individual stress curves as functions of the plastic strain were found 
nearly linear and reversible, where the plastic deformation obeys an effective
equation of motion with a nearly linear force.  For small plastic deformation, 
the means of the stress-strain curves are power law over two decades.  Here
and for somewhat larger plastic deformations, the mean stress-strain curves
converge for larger sizes, while their variances shrink, both indicating the
existence of a thermodynamical limit. The converging averages decrease with
increasing size, in accordance with size-effects from experiments.  For large
plastic deformations, where steady flow sets in, thermodynamical limit was not
realized in this model system.
\end{abstract}


\maketitle

{\em Introduction and overview:} As known from research in the past decade, the stress-strain curve 
of micron-scale specimens contains random steps, where the plateaus mark
avalanches, a behavior also observed in simulations.
\cite{miguel_intermittent_2001, uchic_science_2004,review_uchic,
csikor2007dislocation,ispanovity_acta_mat_2013,ispanovity_submicr_2010}
Recently there has 
been a series of studies about the statistics of
avalanches, and some on probabilistic properties of the yield stress,
\cite{beato_arxiv_2011,csikor2007dislocation,dahmen_avalanche_2012} but the
dependence of these features on the sample size was less investigated.
\cite{beato_arxiv_2011,goldenfeld_jam_2011,derlet2014probabilistic}
The stair like stress-strain response caused by the dislocation avalanches become significant if 
the system size is in the order of 1$\mu$m. Nevertheless, the size dependence of the plastic 
deformation is observed already at much larger system sizes. In most cases the smaller sample 
requires larger stress level to get the same deformation. The effect is traditionally modeled by 
large scale discrete dislocation dynamical (DDD) 
\cite{balint2006,guruprasad2008phenomenological,weygang2008,benzerga2009} simulations, 
phenomenological ``nonlocal continuum theories'' in which an appropriate gradient term is added 
to the stress-strain relation \cite{fleck2001reformulation,gurtin2002gradient}, or by continuum 
theory of dislocations 
\cite{groma2003spatial,groma2006debye,acharya2006,kratochvil2008,mesarovic2010,sandfeld2011,poh2013}
. It is obvious, however, that a continuum theory is not applicable if the dislocation spacing is 
comparable to the system size.        

In this paper the size-dependence of plastic deformations is studied by DDD simulation of 
systems with various numbers of interacting
linear edge dislocations while the stress is slowly increased. Note that elastic
deformations are not considered in this model, so strain and deformation are
understood as purely plastic.  Our main observations in this work  are as
follows. Between avalanches the stress grows close to linearly with the
deformation.  Tests of load cycles show that here the deformation is
approximately reversible, so plasticity appears as a randomly alternating
sequence of quasi-reversible deformations and avalanches.  The linearity
coefficient is random even for systems of the same size (number of dislocations
$N$), with a sharpening distribution for larger $N$.  In each realization,
assuming an effective equation of motion for the deformation, we find a
close-to-linear effective force, with a random spring constant.  The full
individual staircase-like stress-strain curves of the same $N$ also have a mean,
which is a smooth function, different for different $N$'s.  Up to a certain
threshold deformation $\gamma_{\mathrm{th}}$ the mean stress-strain curve seems
to converge for large $N$, while the variance goes to zero. So for large $N$
essentially the same sharp stress-strain curve emerges for each realization,
indicating the existence of a thermodynamic limit. This refines the finding of
Tsekenis et al.\ \cite{goldenfeld_jam_2011}, where the natural scaling by
$\sqrt{N}$ let curves collapse even for smaller $N$. Within the region of
thermodynamical limit $\gamma<\gamma_{\mathrm{th}}$, for two decades in the
strain, the mean stress-strain curve is a power law, with an exponent decreasing
from $1$ for small $N$ to about $0.8$ for large $N$. This is caused by the
alternation of linear, quasi-reversible segments and plateaus of avalanches with
random lengths in the stress-strain response function. Furthermore,  for the
model under study, there seems to be no thermodynamical limit for large
deformations  in the model under study.

{\em Simulation method:} Plastic deformation is mainly due to the motion of
dislocations, \cite{nabarro1967theory} interacting via a long-range stress field.  
Here we study one of the simplest models of dislocation systems described as
follows.\cite{groma_comp_sim_1993, miguel_intermittent_2001, csikor_role_2009, laurson_2010, laurson_2012} 
Straight and parallel edge dislocations with parallel slip axes are considered,
essentially a $2$D cross section of a $3$D system. Periodic boundary
conditions are used on a square of side $L$, the slip axes are parallel to one
edge of the square (the $x$ axis), and for each realization dislocations are 
initially randomly placed with a uniform distribution.  In the 
beginning each realization contains a fixed number $N$ of dislocations, with 
equal number of positive and negative Burgers vectors $s(b,0)$, where $b$ is 
the lattice constant and $s=\pm 1$.  Only dislocation glide is taken into 
account so the dislocations' vertical ($y$) coordinates are constant. 
 
One positive dislocation induces the shear stress field 
\begin{align} 
  \tau_{\text{ind}}(\vec r)= bD x(x^2-y^2) / r^{4}, 
\end{align} 
where $\vec r=(x,y)$ is the radius vector from that dislocation, $r=|\vec
r|$,  $D=\mu/[2\pi(1-\nu)]$, with $\mu$ being the shear modulus and $\nu$ the
Poisson number. The $i$'th dislocation is exposed to the shear stress field of
all the others' $j\neq i$
and to the external field taken to be the uniform $\tau_{\mathrm{ext}}$, wherein
it performs overdamped motion with drag coefficient $B$.  Thus the equation
of motion of the $i$'th dislocation
is \cite{ispanovity_submicr_2010,ispanovity_relax_2011}
\begin{align} \label{eq:mozgasegyenlet}
\dot{x}_i =B^{-1}b s_i \Bigg[\sum_{\begin{subarray}{l}
         j=1 \\
        j\neq i\end{subarray}}^N s_j\tau_{\textrm{ind}}(\vec r_i - \vec r_j)
+\tau_{\textrm{ext}}\Bigg],
\end{align} 
where the $s_k$ is the sign of the $k$'th dislocation and $\vec r_k=(x_k,y_k)$
its position. The plastic strain is calculated by $\gamma=b/L^2\sum_i
s_i\Delta x_i$, where $\Delta x$ is the change in the $x$ coordinate relative
to the initial value. This equation  is rendered periodic numerically by
including sufficiently many mirror images of the $j$'th dislocation.   The
resulting equation of motion is solved with $4.5$th order Runge-Kutta method.
Adaptive step size is used to better treat narrow dipoles. For very narrow
dipoles would demand excessive computation time, we annihilate (different
sign) or merge (same sign) dislocations if they are closer than
$0.05\,L/\sqrt{N}$. Whereas annihilation decreases the dislocation
number, in order to avoid ambiguity,  in the conversion formulas we use the
original dislocation number $N$. Note that dislocations are not created in
our model, corresponding to non-source-controlled plastic deformations. 
Throughout the paper simulations with $N=32,64,128,256,512,1024,2048$ were 
considered with ensembles numbering $10^4, 3000, 2000, 800,300,100,80$,
respectively, and for the largest sizes our computational
power allowed the scanning only of more restriced regions of simulated strain.

Equation \eqref{eq:mozgasegyenlet} is represented in the computer by
$B_{\mathrm{cp}}=D_{\mathrm{cp}}=b_{\mathrm{cp}}=L_{\mathrm{cp}}=1$,
yielding the density $\rho_{\mathrm{cp}}=N/L_{\mathrm{cp}}^2=N$.  The mapping to
different sample sizes, while the physical dislocation density is kept constant,
occurs by our introducing natural quantities $\gamma =
\gamma_{\mathrm{cp}}/\sqrt{N}$, $\tau=\tau_{\mathrm{cp}} /\sqrt{N}$,
$x=x_{\mathrm{cp}} \sqrt{N}$, $t = t_{\mathrm{cp}} {N}$, used throughout this
paper. \cite{csikor_role_2009} Then physical 
quantities are obtained from the physical density $\rho_{\textrm{ph}}$ as
$L_{\textrm{ph}}=\sqrt{N/\rho_{\textrm{ph}}}$, 
$\gamma_{\textrm{ph}}=\gamma b_{\textrm{ph}} \sqrt{\rho_{\textrm{ph}}}$,
$\tau_{\textrm{ph}}=\tau b_{\textrm{ph}} D_{\textrm{ph}}
\sqrt{\rho_{\textrm{ph}}}$, $t_{\textrm{ph}}= t B_{\textrm{ph}}
(b_{\textrm{ph}}^2  D_{\textrm{ph}} \rho_{\textrm{ph}})$.
To give an example, if on the natural scale $\gamma=0.35$ then for the generic
physical variables $\rho_{\textrm{ph}}=2 \cdot 10^{14}\textrm{m}^{-2}$ and
$b_{\textrm{ph}}=2 \textrm{\AA}$ we would obtain the physical deformation
$\gamma_{\textrm{ph}}=0.1\%$. 

In the scenario presented here firstly we let the system relax without external
stress to form the initial state with $\gamma=0$. Then we apply quasi-static
stress loading, i.e., we increase the
external stress by a small rate, similar to what was applied
earlier.\cite{beato_arxiv_2011,goldenfeld_jam_2011,derlet2013}  Having
experimented with various stress rates we finally chose
$\dot\tau_{\mathrm{ext}}=5\cdot 10^{-5}$. The stress is kept increasing so long
as the mean absolute velocity of the dislocations remains under the threshold
$5\cdot 10^{-4}$.  If that threshold is surpassed (this is our definition of an avalanche)
then the external stress is kept constant until the mean absolute velocity drops
again below the threshold. The end state is a
steady flow, that is, an infinite avalanche, because in the absence of
dislocation creation no work hardening takes place for large deformations.

\begin{figure}[t!bh]
\centering
\includegraphics[angle=0,width=8.6cm]{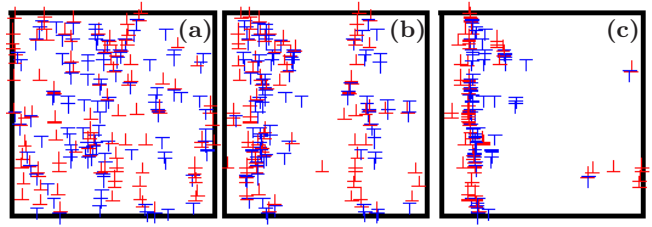}
\begin{picture}(0,0)
 \put(-57.5,82){\textbf{(a)}}
 \put(22.5,82){\textbf{(b)}}
 \put(104.5,82){\textbf{(c)}}
\end{picture}
\caption{(Color online) Typical configuration at (a) relaxed state without
external stress ($\gamma=0$), (b) relaxed in the presence of medium external
stress ($\gamma\approx 1$), (c) with external stress just
below the steady flow ($\gamma\approx 1000$). The $\bot$ (red) and the $\top$
(blue) denote the $s=+/-$ signs, respectively.}\label{fig:confs}
\end{figure}

{\em Individual stress-strain curves and effective motion between
avalanches:}
We first show typical dislocation configurations for various plastic
deformations
$\gamma$ along our scenario in Fig.\ \ref{fig:confs}. 
Firstly, dislocations relax in the absence of
external stress, forming a random-looking configuration of numerous smaller
clusters, see Fig.\ \ref{fig:confs}a. Due to the increasing external stress
the  deformation $\gamma$ generically increases, while clusters grow mainly in
the $y$ direction (\ref{fig:confs}b).  Beyond some threshold steady flow
emerges, marked usually by  a single dipolar wall, spanning across the whole
simulation area (\ref{fig:confs}c), while one-or-two dislocations are circling
quickly along their slip axes, in conformance to the periodic boundary
conditions. Hence we must conclude that in the steady flow boundary effects are
important, thus our model may not be realistic in this region, and
so we concentrate our study to smaller deformations.  
 
\begin{figure}[t!hb]
\centering
\includegraphics[angle=270,width=8.6cm]{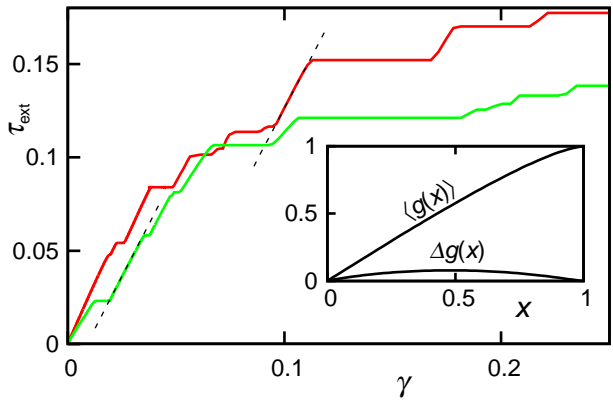}
\caption{(Color online) Stress-strain curves $\tau_{\mathrm{ext}}(\gamma)$ of two realizations
of $N=512$. Segments between plateaus are found to be nearly linear, dashed lines
are guide to the eye. Inset: averages of increasing segments normalized as in
Eq.\ \eqref{eq:gfuggveny} for fixed $N$'s all fall on the same curve, close to a
linear function.  The lower arc is the standard deviation, again
nearly independent on $N$.}
\label{fig:lefolyas} 
\end{figure}

The  plastic stress-strain curves  $\tau_{\rm ext}(\gamma)$ of individual
realizations 
are like staircases, they exhibit a sequence of plateaus of constant stress
corresponding to avalanches, in accordance with earlier results,
\cite{miguel_intermittent_2001, uchic_science_2004,review_uchic,
csikor2007dislocation,ispanovity_acta_mat_2013,beato_arxiv_2011}   see Fig.\
\ref{fig:lefolyas}.  A novel
observation here, to our knowledge not noted earlier, is that between plateaus,
where the stress increases,  the $\tau_{\rm ext}(\gamma)$ functions are nearly linear.
For the sake of comparison we normalized each increasing segment to a function
$g(x)$ starting and ending at the opposite corners of the unit square as
\begin{align} \label{eq:gfuggveny}
g(x)= \frac{\tau_{\mathrm{ext}}\left(x(\gamma_{1}-\gamma_{0})+
\gamma_{0}\right)-\tau_{0}}{\tau_{1}-\tau_{0}}, \ \ x\in [0,1],
\end{align} where $\gamma_0$, $\gamma_1$ and $\tau_0$, $\tau_1$ mark the borders
of the chosen segment. Such normalized segments of individual runs with
$\gamma<0.2$ were separately averaged for fixed sizes $N$, and the resulting
curves fall onto each other and form $\langle g(x)\rangle$ in the inset of Fig.\
\ref{fig:lefolyas}. Here, also the standard deviation of the normalized segments
is plotted, which is again nearly the same for various $N$'s. So the segments
between avalanches on the stress-strain curves, normalized according to Eq.\
\eqref{eq:gfuggveny},
follow on the average a universal, nearly linear form, with a universal
variance. 
 
In order to test properties of the close-to-linear segments, we have run a few
loading cycles on individual realizations with various stress rates. In most of
the cases, apart from a small, smooth, transient due to the finiteness of the
stress rate, and from tiny avalanches, reversibility was found. 
Motivated by this ``quasi-reversible'' response, we surmise that deformations
obey an effective equation of motion
\begin{align} \label{eq:effektiv-mozgegy}
\dot{\gamma}(t) \approx  -F(\gamma(t)-\gamma_0)+\tau_{\mathrm{ext}}(t)-\tau_0,
 \end{align}
where $(\gamma_0,\tau_0)$ the endpoint of the last avalanche, where the system 
is assumed to be in equilibrium, and $F$ is the effective restoring force, 
depending only on the increment $\gamma-\gamma_0$. Similarly as 
\eqref{eq:gfuggveny} associates $g(x)$ with the stress, we normalize the force 
$F(\gamma-\gamma_0)$ onto the unit square. Again, like the external stress in
the inset of Fig.\ \ref{fig:lefolyas}, the mean normalized force very weakly
depends on $N$, and this universal function is astonishingly close to $g(x)$,
which is the consequence of the low stress rate resulting
$\dot{\gamma}(\gamma)\ll\tau_{\mathrm{ext}}(\gamma)$. Thus even the force is
very close to linear. We emphasize, that near equilibrium of a
dislocation configuration, for small displacements, the elastic energy of course grows
quadratically, so there the response
should be linear. The remarkable feature in our case is that linearity holds way
up to near the onset of the next avalanche.  This instability is indeed marked
by the slight curving of the universal function close to one in Fig.\
\ref{fig:lefolyas}. Note that its slope there is not zero, whereas it would
be zero in the case of a force-activated escape from a 1D potential, because of
rare negative avalanches with  $\dot\gamma<0$.  
 
\begin{figure}[htb]
\centering
\includegraphics[angle=270,width=8.6cm]{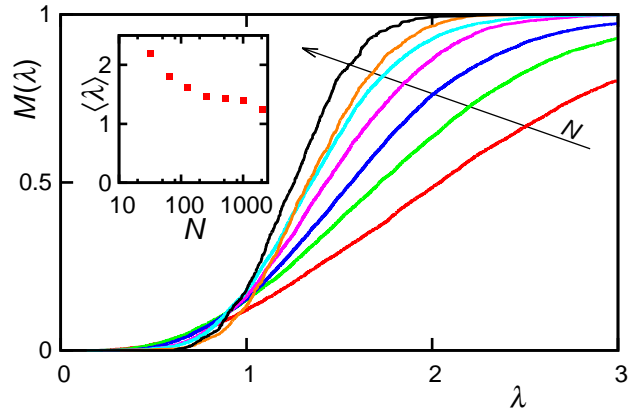}
\caption{(Color online) Cumulative probability distribution function 
$M(\lambda,N)$ of the steepness $\lambda$ of the quasi-linear segments with
system sizes $N=32, 64,\dots ,2048$ for $\gamma \le 0.2$. The arrow points
towards increasing $N$ (colors distinguish sizes). Inset: the mean
$\langle\lambda\rangle$ as functions of the system size $N$. }
\label{fig:tauslope-distr}
\end{figure}

In order to visualize the statistical nature of the quasi-reversible regions,
in Fig.\ \ref{fig:tauslope-distr} we plot the cumulative probability
distribution function $M(\lambda)$ of the steepness 
$\lambda=(\tau_{1}-\tau_{0})/(\gamma_1- \gamma_0)$ for $\gamma \le 0.2$
(chosen as a practical value). The curves visibly contract with increasing $N$,
showing convergence to a finite mean.
A nearly linear response here  means that, due to the interaction of 
dislocations, an effective shear modulus arises. The latter can be interpreted 
as the plastic component of the total empirical shear modulus in real crystals. 
Note that in the present model elastic deformations are not included. 
 
\begin{figure}[t!]
\centering
\includegraphics[angle=270,width=8.6cm]{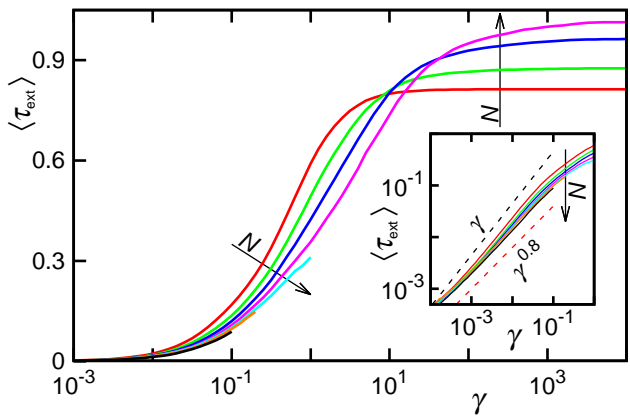}
\caption{(Color online) Mean stress $\langle\tau_{\mathrm{ext}}\rangle$ as a
function of
strain, for system sizes $N=32, 64,\dots ,2048$, for large $N$ only smaller
$\gamma$'s were considered.  Inset: log-log plot demonstrates 
the power law for $5\cdot10^{-4}\lesssim  \gamma \lesssim 0.05$. The arrows
point towards increasing $N$ (colors distinguish sizes). The two dashed lines are
guide to the eye.}
\label{fig:tauatlag} 
\end{figure}

{\em Mean stress as function of the strain:} 
So far we concentrated on the quasi-reversible segments between avalanches, now 
we turn to the global statistical behavior of  stress-strain curves.  Firstly, 
we plot the average $\langle\tau_{\mathrm{ext}}\rangle(\gamma,N)$ over ensembles
with fixed $N$ in Fig.\ \ref{fig:tauatlag}.  A main feature is the power law
behavior over decades up to a strain approximately $0.05$, (see inset), with
exponent close to $1$ for 
small sizes and decreasing with size to about $0.8$.  We can interpret this 
feature such that the nearly linear segments of the stress-strain curves are 
interrupted by the avalanche plateaus just in the way that an effective 
power function with a smaller-than-one exponent emerges. That is, avalanches 
soften the linearity of quasi-reversible segments and give rise statistically 
to a power law.  

On physical scales, taking a lattice constant $b=2 \textrm{\AA}$, a
dislocation density $\rho=2\cdot 10^{14}\mathrm{m}^{-2}$, the $\gamma=0.05$
corresponds to $\gamma_{\mathrm{ph}}\approx 0.015\%$.  It is important to note
that this
value is much smaller than the $\gamma_{\mathrm{ph}}=0.2\%$ threshold value
customarily considered to be the yield strain in engineering practice. The fact
that the power law arises way below the empirical yield value indicates the
determining role of avalanches for much lower plastic strains than expected
earlier. On the other hand, this is in a convincing agreement with fatigue
experiments on single crystals that showed irreversibility already below the
physical range of $\gamma_{\mathrm{ph}}=0.01\%$.
\cite{mughrabi_cyclic_slip_irrev} 
 
Another important property seen in Fig.\ \ref{fig:tauatlag} is that for
$\gamma\lesssim 1$ the stress-strain curves seem to converge for large $N$, a
criterion for the existence of a thermodynamical limit.  On the other hand,
for large strains, the $N\to\infty$ tendency is inconclusive from our
simulation, the stress values may even diverge.
For intermediate strains $1\lesssim \gamma\lesssim 100$ the convergent bundle of
the curves switches order for the sizes we considered. Note that for fixed
strains $\gamma\lesssim 1$ the stresses decrease with $N$. As discussed earlier
in connection with physical units, increasing $N$ here can mean increasing size
with constant dislocation density. Therefore, larger stresses for smaller $N$'s
as in Fig.\ \ref{fig:tauatlag} can be interpreted as a version of the property
``smaller is harder''. We emphasize, however, that our model has periodic
boundary conditions, thus pileups, commonly held responsible for this
phenomenon,\cite{kovacs_zsoldos} cannot develop.  This tendency reverses for
large strains, where the stress increases with $N$.
In this region, however, where the configuration resembling a single wall
forms, as seen in Fig.\ \ref{fig:confs}c, we do not suggest the model
bears general relevance to real materials.

\begin{figure}[t!]
\centering
\includegraphics[angle=270,width=8.6cm]{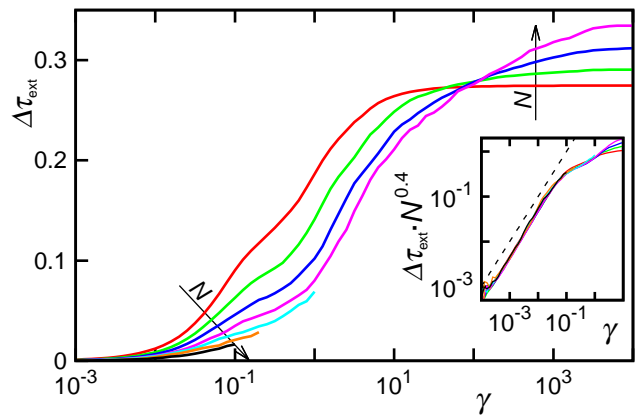}
\caption{(Color online) Standard deviation 
$\Delta\tau_{\mathrm{ext}}$ of the stress as function of the strain
$\gamma$, for system 
sizes $N=32,64,\dots,2048$, for large $N$ only smaller $\gamma$'s were 
considered. The arrows point towards increasing $N$ (colors distinguish sizes).
Inset: log-log plot of $\Delta\tau_{\mathrm{ext}}\cdot N^{0.4}$ shows
collapse, the dashed line with power unity is a guide to the eye.}
\label{fig:tauszoras} 
\end{figure}

{\em Standard deviation of the stress as a function of strain:}
Given the fact that the stress-strain response for macroscopic crystals with a
fixed orientation is a well-defined, sharp curve, it is expected that the
variance vanishes with increasing size. Accordingly,  a decreasing variance was
observed in micropillar experiments by Uchic {\em et al}.
\cite{uchic_science_2004}  To study this effect, we plotted in Fig.\
\ref{fig:tauszoras} the standard deviation $\Delta\tau_{\mathrm{ext}}(\gamma,N)$
of the stress for ensembles with fixed $N$ for different strains. In the region
$\gamma\lesssim 1$, where the mean stress 
converged with $N$  (see Fig.\ \ref{fig:tauatlag}), the standard deviation 
decreases.  We tested a power law convergence by plotting  
$\Delta\tau_{\mathrm{ext}}(\gamma,N)\cdot N^{0.4}$ in the inset of Fig.\
\ref{fig:tauszoras}, and indeed the collapse demonstrates that the deviation
vanishes like $1/N^{0.4}$.  Thus, recalling that we found a sharp limit
for the average stress when $\gamma\lesssim 1$, we can conclude that in this
region there is a thermodynamical limit.   On the contrary, for large strains
Fig.\ \ref{fig:tauszoras} does not show convergence of the 
standard deviation, like there was no visible convergence of the mean in Fig.\ 
\ref{fig:tauatlag} either, so here thermodynamical limit is absent.

{\em Conclusion and outlook:} In this paper our aim was twofold.  On the one
hand, we uncovered quasi-reversible behavior with nearly linear response between
avalanches in a 2D model of simulated dislocations.  For large sizes, the
effective plastic shear modulus appears to converge in average.  As to the full 
stress-strain curves, the quasi-reversible segments conspire with the avalanche 
plateaus to yield on the average a power response curve.  The second main
question was about the thermodynamical limit, which is achieved by both a
convergent mean and a vanishing variance of the stress for $\gamma \lesssim 1$.
 In this region we observed also the analog of the size-effect, as found in
micropillars.\cite{uchic_science_2004}  For largest strains
thermodynamical limit is not reached and we do not consider our simulations as
conclusive there. 
 
Our study opens a series of questions. The results on the quasi-reversible 
behavior between avalanches call for more detailed investigations.  The 
statistical properties of the finite-size behavior is best characterized by 
distribution functions, among which here we only described that of the local 
effective plastic shear modulus, characterizing quasi-reversible regions. The 
distribution of the stresses is of obvious interest, and at avalanches are 
expected to be related to extreme statistics.  A longstanding problem in this 
area is the transition to steady flow, wherein the absence of a single 
critical point, rather critical behavior for all strains, has been shown 
before,\cite{ispanovity_relax_2011} but a detailed study is still overdue. 
Carrying forth experiments on micropillars\cite{ispanovity_acta_mat_2013} 
with various sizes and their comparison to the prediction from simulations 
would be of immediate interest. 

Financial supports of the Hungarian Scientific Research Fund (OTKA) under 
contract numbers K-105335 and PD-105256, and of the European Commission under
grant agreement No. CIG-321842 (StochPlast) are acknowledged. 

\bibliography{deform_quasilin_aval_paper}

\end{document}